\documentclass[a4paper,twocolumn,superscriptaddress,preprintnumbers,amsmath,amssymb,floatfix]{revtex4}
\usepackage{graphicx}
\usepackage{textcomp}					

\begin{document}

\title{Minimal bundling of single-walled carbon nanotubes comprising\\ vertically aligned films}



\author{Erik Einarsson}
\affiliation{Dept.~of Mechanical Engineering, The University of Tokyo, Tokyo 113-8656, Japan}
\author{Hidetsugu Shiozawa}
\affiliation{Inst.~for Solid State and Materials Research (IFW) Dresden, D-01171 Dresden, Germany}
\author{Christian Kramberger}
\affiliation{Inst.~for Solid State and Materials Research (IFW) Dresden, D-01171 Dresden, Germany}
\author{Mark H.~R\"{u}mmeli}
\affiliation{Inst.~for Solid State and Materials Research (IFW) Dresden, D-01171 Dresden, Germany}
\author{Alex Gr\"{u}neis}
\affiliation{Inst.~for Solid State and Materials Research (IFW) Dresden, D-01171 Dresden, Germany}
\author{Thomas Pichler}
\affiliation{Inst.~for Solid State and Materials Research (IFW) Dresden, D-01171 Dresden, Germany}
\author{Shigeo Maruyama}
\email{maruyama@photon.t.u-tokyo.ac.jp}
\affiliation{Dept.~of Mechanical Engineering, The University of Tokyo, Tokyo 113-8656, Japan}

\date{\today}

\begin{abstract}
A freestanding film of vertically aligned single-walled carbon nanotubes (VA-SWNTs) synthesized by the alcohol catalytic chemical vapor deposition (ACCVD) method was observed directly by transmission electron microscopy (TEM). These observations revealed that the film is comprised primarily of small SWNT bundles, typically containing 3-8 SWNTs. The lack of significant bundling is supported by electron diffraction spectra, in which no bundle peak is observed, and by electron energy-loss spectroscopy (EELS), which shows minimal dielectric screening. This indicates the electrical properties of the VA-SWNT film are dominated by the one-dimensional nature of SWNTs, rather than behaving as a bulk material.
\end{abstract}

\maketitle

\section{Introduction}
Many potential applications of single-walled carbon nanotubes (SWNTs) have been proposed since their discovery,\cite{Iijima-Nature_SWNT} most of which are based on exploiting the novel physical properties of SWNTs arising from their quasi-one-dimensional structure.\cite{Saito-text,Yao-Avouris-transport_review,Dresselhaus-text} This one-dimensionality induces strong anisotropy in many SWNT properties, such as optical absorption,\cite{Hwang-polarized_spectroscopy-aligned_SWNTs,Ichida-anisotropic_optical_polymer,Murakami-PRL94,Islam-SWNT_absorption} thermal conductivity,\cite{Ruoff-mech_therm_prop,Fischer-mag_align_transport} etc., hence, controlling the orientation of SWNTs is highly desirable. Vertically aligned (VA-) synthesis of SWNTs was first realized11 using a dip-coat catalyst preparation method,\cite{Murakami-CPL377} and were synthesized from alcohol by ACCVD.\cite{Maruyama-CPL360} Since this first report, vertically aligned growth has been realized by several other groups using various synthesis techniques such as water-assisted CVD,\cite{Hata-supergrowth} oxygen-assisted CVD,\cite{Zhang-Dai-roles_of_H2_and_O2} point-arc microwave plasma CVD,\cite{Zhong-VASWNTs-plasmaCVD} molecular-beam CVD,\cite{Eres-molecular_beam_VASWNT_growth} and hot-filament CVD.\cite{Xu-Hauge-hot_filament_VASWNTs} It has also been shown that the catalyst can be optimized to obtain VA-SWNT growth by traditional CVD methods as well.\cite{Maruyama-CPL403,Noda-Co-Mo_combinatorial,Zhang-Resasco-catal_denisty}

In addition to controlling the orientation of SWNTs, there is another challenge regarding SWNT synthesis. Many of the unique 1D properties of SWNTs are lost due to bundling of the nanotubes, which not only makes many proposed applications impractical or impossible, but makes measurements of some fundamental physical properties of SWNTs all the more challenging. In particular, the effects of bundling on the electronic structure of SWNTs have been addressed in many reports in the literature.\cite{Delaney-pseudogaps_CNTs-nature,Ouyang-Leiber-nature,Reich-PRB_band_structure,Maarouf-electronic_structure_CNT_ropes,Tomanek-electron_structure_CNT_ropes} 

In this paper, we report the first transmission electron microscope (TEM) observations of freestanding, vertically aligned (VA-) SWNT films produced from alcohol, which reveal the films are comprised of very small SWNT bundles. This result is supported by electron diffraction and electron energy-loss spectroscopy (EELS), the latter also indicating the small bundles significantly impact the electronic properties of the VA-SWNT films. 

\section{Experimental Methods}
\subsection{Preparation of vertically aligned SWNTs by the ACCVD method}
\begin{figure}[tb]
	\includegraphics[width=0.45\textwidth]{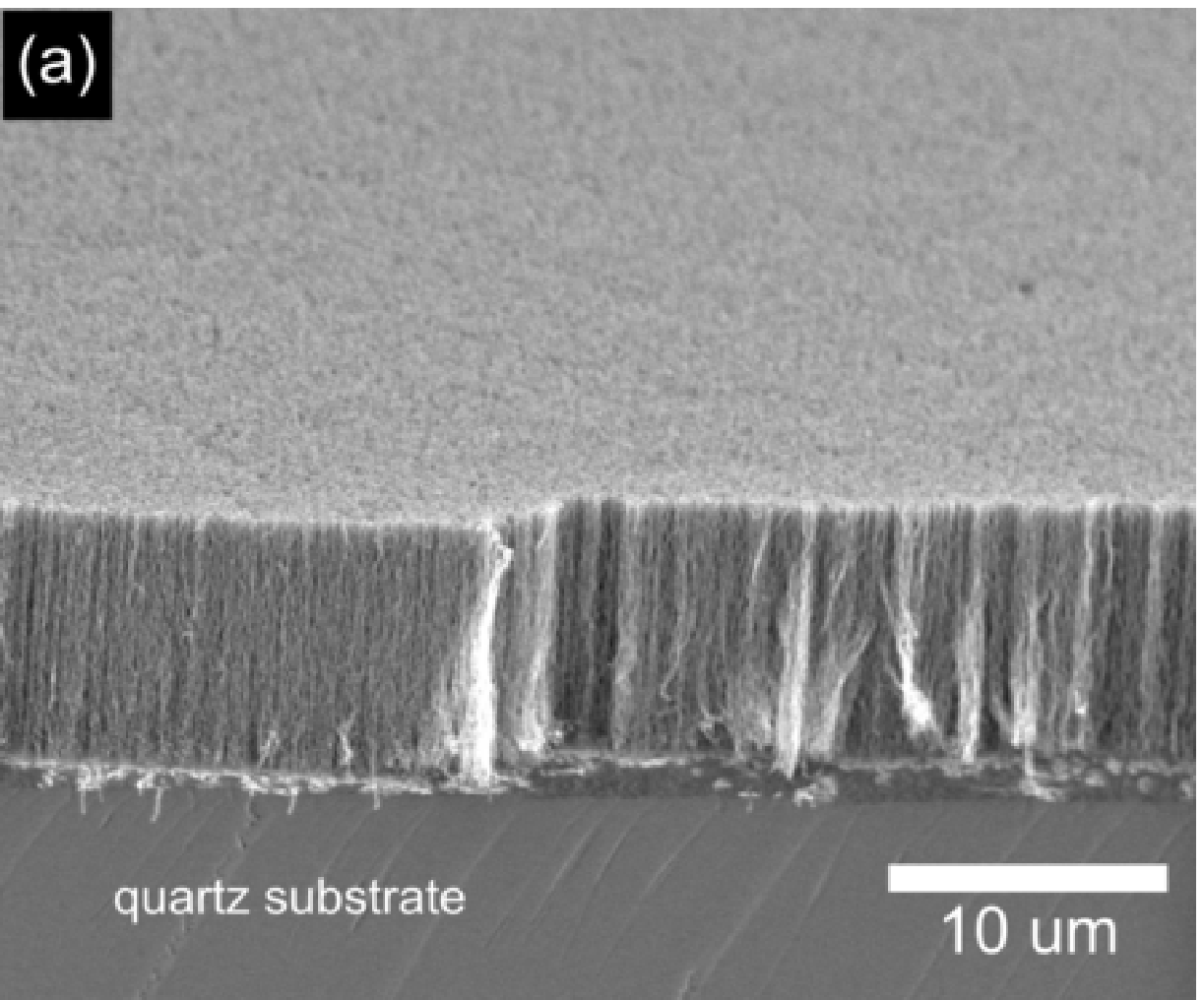}
	\includegraphics[width=0.45\textwidth]{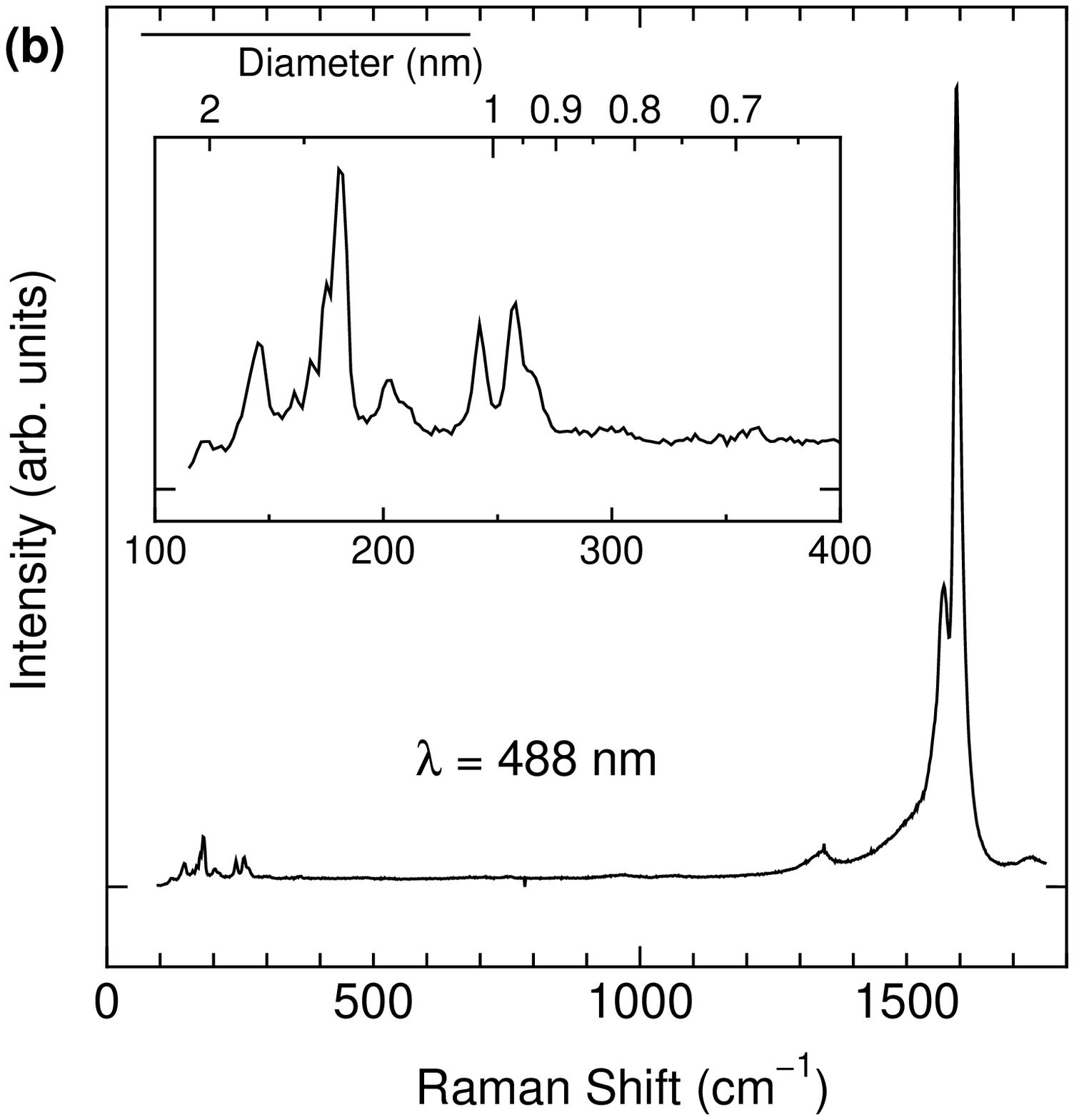}
	\caption{\label{VASWNTs_Raman}SEM image of vertically aligned SWNTs synthesized from alcohol. The film thickness can be controlled from 1 to more than 30\,$\mu$m.\cite{Einarsson-unpub} (b) A Raman spectrum representative of VA-SWNTs produced by the alcohol CVD method.}
\end{figure}

The VA-SWNTs used in this study were synthesized from ethanol by the alcohol catalytic chemical vapor deposition (ACCVD) method.\cite{Murakami-CPL377,Maruyama-CPL360} This method has been described in detail in previous reports,\cite{Murakami-CPL385,Maruyama-CPL403} but the general process is as follows. Catalyst is prepared by dip-coating optically polished quartz glass into a solution of Co and Mo acetate dissolved in ethanol (metal content 0.01\,wt\% each). The substrate is then baked in air at 400\,\textdegree\/C for 5 minutes, which oxidizes the deposited metals and prevents agglomeration at higher temperatures. Immediately prior to CVD, the sample is heated to the growth temperature (typically 800 \textdegree\/C) under 40\,kPa of flowing Ar/H$_{2}$ (3\% H$_{2}$, Ar balance, flow rate of 300\,sccm). After reaching the growth temperature, the Ar/H$_{2}$ flow is stopped, and ethanol is introduced into the chamber for 10-15 minutes at a pressure of 0.8\,kPa. An SEM image of VA-SWNTs produced by this method is shown in Fig.~\ref{VASWNTs_Raman}\,a. Figure \ref{VASWNTs_Raman}\,b is a corresponding resonance Raman spectrum, which reveals the presence of high-purity SWNTs.\cite{Dresselhaus-Eklund-Phonons} The dominant RBM peak at 180\,cm$^{-1}$ (see inset) has been shown to be indicative of vertical alignment.\cite{Murakami-PRB71} During ACCVD synthesis, the VA-SWNT film thickness was monitored using an in situ optical absorbance technique,\cite{Maruyama-CPL403,Einarsson-unpub} allowing accurate control over the film thickness up to $\sim$30\,$\mu$m. 

\subsection{Transfer onto TEM grid}
\begin{figure}[tb]
	\includegraphics[width=0.45\textwidth]{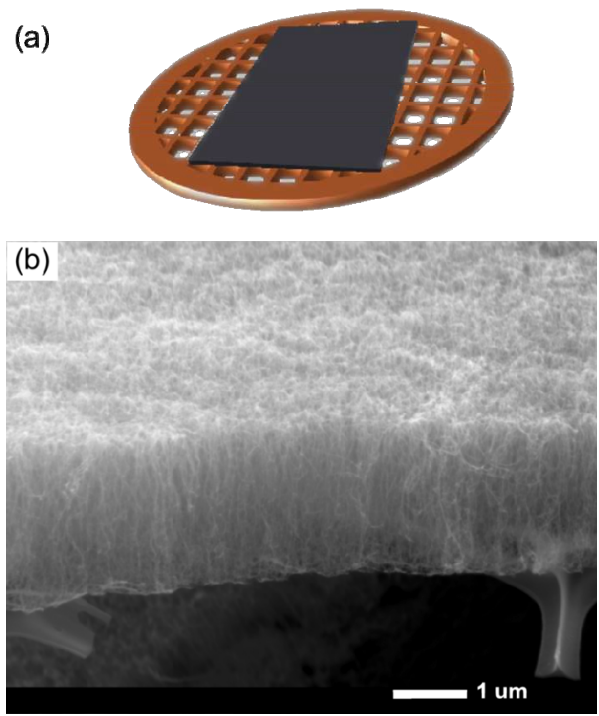}
	\caption{\label{VASWNTs_on_TEM_grid}(a) illustration of and (b) FE-SEM image of a vertically aligned SWNT film sitting atop a TEM grid.}
\end{figure}

After synthesis, the VA-SWNT film was removed from the quartz substrate on which it was grown, and transferred onto a TEM grid using the hot-water assisted process described in Ref.~\cite{Murakami-CPL422}. Since this method preserves vertical alignment, the resulting sample was a freestanding VA-SWNT film sitting atop a TEM grid, as shown by the SEM image in Fig.~\ref{VASWNTs_on_TEM_grid}\,a. After transfer onto the TEM grid, electron energy-loss spectroscopy and electron diffraction measurements were performed at room temperature under ultra-high vacuum using a purpose-built high-resolution spectrometer having good energy and momentum resolution, as described in Ref.~\cite{Fink-Adv_Electron_phys}. To ensure electron transparency, thin films of 2 and 7\,$\mu$m were used in the experiments presented here.

\section{Results}
\subsection{Electron diffraction and EELS}
\begin{figure}[tb]
	\includegraphics[width=0.45\textwidth]{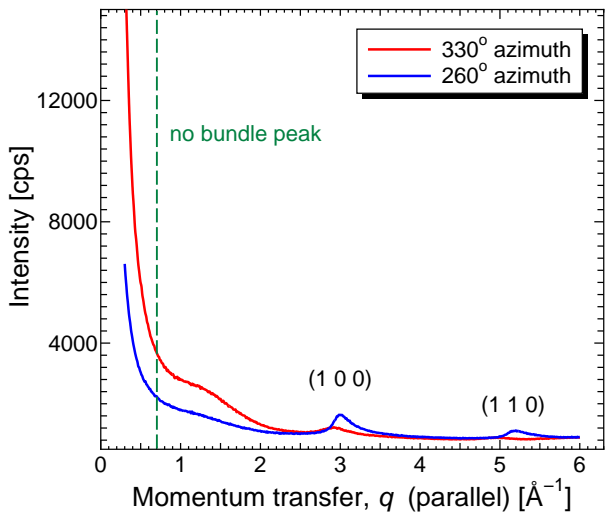}
	\caption{\label{electron_diffraction}Electron diffraction intensity profiles of the VA-SWNT film showing alignment-induced anisotropy, but no bundle peak in the low-$q$ region.}
\end{figure}

Electron diffraction spectra obtained from SWNT samples have three characteristic peaks, including the (1\,1\,0) and (1\,0\,0) peaks from the graphitic structure, and a large peak at low momentum transfer (near 0.7\,\AA$^{-1}$), which is caused by inter-tube scattering within SWNT bundles.\cite{Liu_Pichler-Mag_aligned_SWNTs,Lambin-Carbon-measuring_chirality} Due to the alignment of the VA-SWNTs, we expect an orientation-dependent anisotropy in the electron scattering spectra, shown in Fig.~\ref{electron_diffraction}. As expected, the (1\,0\,0) and (1\,1\,0) reflections (at 2.9 and 5\,\AA$^{-1}$, respectively) are stronger for incidence along the orientation axis,\cite{Lucas-UV_abs_cross-section} however there is no indication of the bundle peak expected at low momentum-transfer. The lack of this bundle peak does not imply all the SWNTs are isolated, but it does indicate that the extent of bundling of SWNTs within the film is minimal. The broad hump between 1 and 2\,\AA$^{-1}$ may be due to the broad diameter distribution of the SWNTs present in the sample.\cite{Murakami-CPL385} 

\begin{figure}[tb]
	\includegraphics[width=0.45\textwidth]{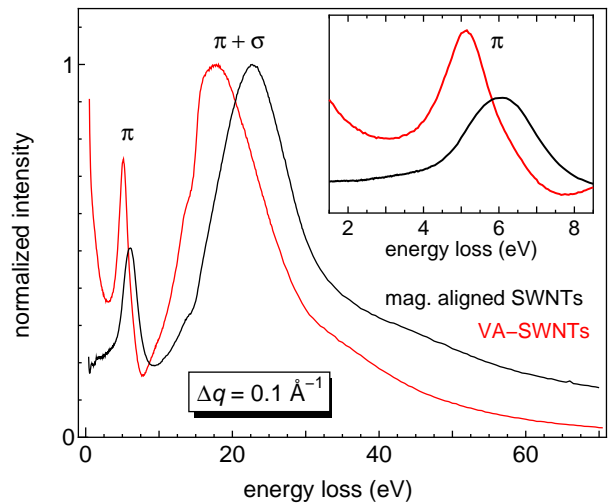}
	\caption{\label{EELS}EELS spectra for momentum transfer $\Delta q$ = 0.1\,\AA$^{-1}$ showing the $\pi$ and $\pi$\,+\,$\sigma$ plasmon peaks for magnetically aligned SWNT bundles (black line, from Ref.~\cite{Liu_Pichler-Mag_aligned_SWNTs}) and VA-SWNTs (red line). The lower energy of the $\pi$ plasmon for VA-SWNTs (see inset) indicated reduced tube-tube interactions.}
\end{figure}

High resolution EELS measurements also indicate minimal SWNT bundling within the VA-SWNT films. Figure \ref{EELS}\,a shows two EELS spectra for momentum transfer $\Delta q$ = 0.1\,\AA$^{-1}$. The red line corresponds to the VA-SWNTs discussed in this report, and the black line corresponds to magnetically aligned bundles of SWNTs.\cite{Liu_Pichler-Mag_aligned_SWNTs} A full analysis of the EELS spectra for these two samples can be found in Ref.~\cite{Kramberger-IWEPNM2007}, but here we focus only on the low momentum transfer region ($\Delta q$ = 0.1\,\AA$^{-1}$), which can be compared to optical absorption spectroscopy. The peaks in these spectra shown in Fig.~\ref{EELS} correspond to plasma oscillations (plasmons)\cite{Lucas-UV_abs_cross-section,Shyu-Lin-loss-spectra_graphite_systems,Pichler-SWNT_plasmons,Knupfer-Carbon-SWNT_EELS} in the $\pi$ electron system (between 5 and 7\,eV) and the combined oscillation of the $\pi$ and $\sigma$ electrons (around 20\,eV). For the VA-SWNTs discussed here, both plasmon excitations are found at lower energies than in the magnetically aligned SWNTs.\cite{Liu_Pichler-Mag_aligned_SWNTs} The $\pi$ plasmon energy is particularly sensitive to the local dielectric response function, which changes with polarization and shielding effects. This peak for the VA-SWNTs (see Fig.~\ref{EELS} inset) appears at 5\,eV, which is a full 1\,eV lower than in the magnetically aligned SWNT sample, and in good agreement with the value of 4.5\,eV previously obtained from polarized optical absorption measurements of VA-SWNTs.\cite{Murakami-PRL94} Considering the electron diffraction results and the sensitivity of the $\pi$ plasmon energy to local shielding effects, this reduced plasmon energy is very likely due to reduced bundling of the SWNTs within the vertically aligned SWNT film.

\subsection{TEM observation}
\begin{figure}[tb!]
	\includegraphics[width=0.40\textwidth]{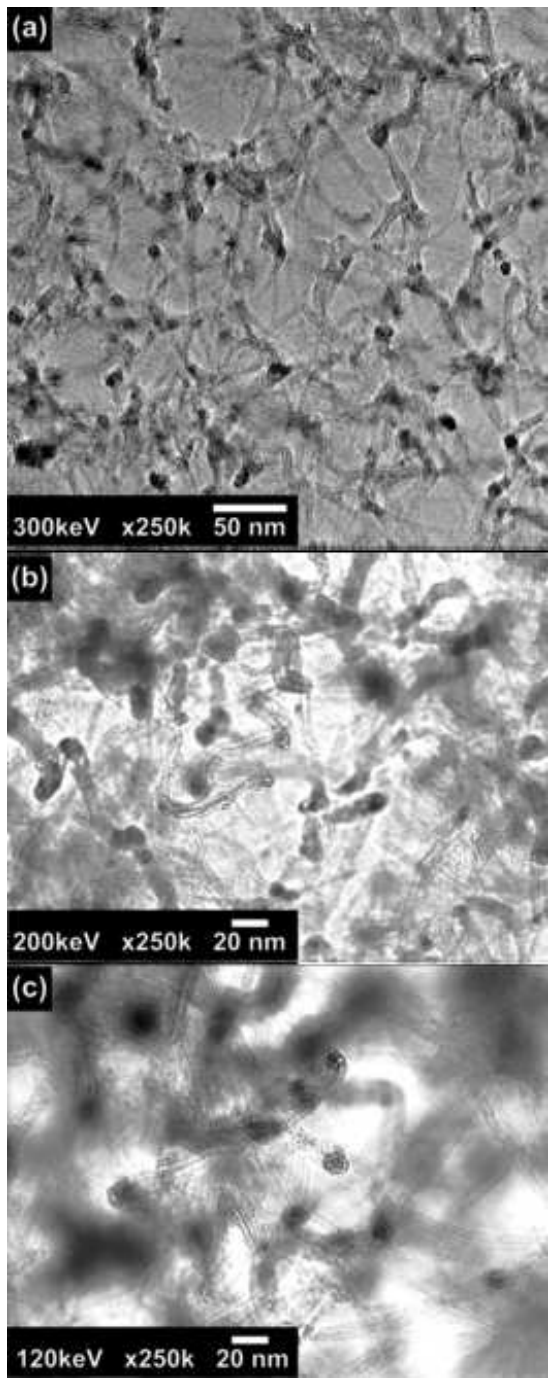}
	\caption{\label{TEM}TEM images along the alignment direction of a freestanding VA-SWNT film. The cross-sections of many small SWNT bundles can be seen. The image in (a) was taken at 300\,keV with a Tecnai F30, while (b) and (c) were taken at 200 and 120\,kV, respectively using a JEOL JEM-2000EX.}
\end{figure}

In order to test this small-bundle hypothesis, the VA-SWNTs were observed in a transmission electron microscope (TEM), such that the perspective was along the alignment direction (i.e. down from the top of the film shown in Fig.~\ref{VASWNTs_on_TEM_grid}\,a). Fig.~\ref{TEM}\,a was taken with an accelerating voltage of 300\,kV using an FEI Tecnai F-30 microscope. Despite the thickness of the sample (2\,$\mu$m), its low density makes it transparent to the electron beam. The high acceleration voltage results in very easy transmission through the sample, thus SWNTs in the foreground and background are not apparent in the image. The dark spots dotting the image are not metal particles or carbonaceous impurities, but are cross-sections of small SWNT bundles passing through the image plane. The large number of cross-sections visible in this image is due to the overall alignment of the SWNTs. The bundles have a diameter of $\leq$ 10 nm, and typically contain 3-8 SWNTs. A few isolated SWNTs were also found dispersed throughout the film.

The images shown in Fig.~\ref{TEM}\,b and \ref{TEM}\,c were taken using reduced acceleration voltages (200 and 120\,kV, respectively. These images were also obtained using a thicker VA-SWNT film (approximately 7\,$\mu$m). The lower acceleration voltage and thicker VA-SWNT film made the thickness of the film much more apparent, particularly in Fig.~\ref{TEM}\,c. The SWNT bundles in the thicker film may be slightly larger than those found in Fig.~\ref{TEM}\,a, tending closer to 6-10 SWNTs per bundle, but the overall internal structure observed in both cases is basically the same. This new perspective on the internal structure supports the EELS and electron diffraction results indicating minimal bundling of the VA-SWNTs. From the TEM images, however, the SWNTs seem less-well aligned than indicated by typical SEM observation. It is possible that the exposed SWNTs at a free surface of the VA-SWNT film differ from the inside, which might explain the difference between SEM and TEM images.

The TEM observations shown above indicate the bundles comprising the VA-SWNT film typically contain 3-10 SWNTs. For bundles of this size, most or all of the SWNTs are on the 'surface' of the bundle, and essentially none in the interior, or bulk of the bundle. Hence, one expects very little change in the dielectric response function caused by surrounding SWNTs. Early calculations by Lin {\em et al}.,\cite{Lin-Chuu-pi_plasmons_CNT_bundles} show the dispersion of the $\pi$ plasmon in individual SWNTs should be larger than in bundles, in agreement with our EELS spectra. A detailed analysis of the EELS results will be addressed in a separate report,\cite{Pichler-EELS_unpub} but these initial results indicate the SWNTs in the vertically aligned film are sufficiently unbundled to retain the 1D electrical nature of the component SWNTs.

\section{Conclusions}
In summary, we report the first TEM observations of a vertically aligned (VA-)SWNT film along the alignment direction. These observations reveal that the VA-SWNTs form unexpectedly small bundles (containing fewer than 10 SWNTs). Electron diffraction and electron energy-loss spectra obtained from these VA-SWNTs show few signs of bundling, including the lack of a bundle peak in the diffraction spectra, and a $\pi$ plasmon energy significantly lower than that of magnetically aligned SWNT bundles.\cite{Liu_Pichler-Mag_aligned_SWNTs} These findings indicate that the VA-SWNT film is composed of sufficiently small bundles that the overall electronic properties of the film are dominated by individual SWNTs rather than bulk SWNT bundles,\cite{Pichler-EELS_unpub} and are very interesting for future studies on the 1D properties of SWNTs, and promising with regard to applications that take advantage of the unique electronic/optical properties of aligned SWNTs.

\begin{acknowledgments}
The authors thank H. Tsunakawa at the University of Tokyo for assistance with TEM measurements.
\end{acknowledgments}

\bibliographystyle{elsart-num}
\bibliography{refs}

\end{document}